# Observation of $h_1(1380)$ in the $J/\psi \to \eta' K\bar{K}\pi$ decay

M. Ablikim,[1] M. N. Achasov,[9,d] S. Ahmed,[14] M. Albrecht,[4] A. Amoroso,[53a,53c] F. F. An,[1] Q. An,[50,40] J. Z. Bai,[1] Y. Bai,[39] O. Bakina,[24] R. Baldini Ferroli,[20a] Y. Ban,[32] D. W. Bennett,[19] J. V. Bennett,[5] N. Berger,[23] M. Bertani,[20a] D. Bettoni,[21a] J. M. Bian,[47] F. Bianchi,[53a,53c] E. Boger,[24,b] I. Boyko,[24] R. A. Briere,[5] H. Cai,[55] X. Cai,[1,40] O. Cakir,[43a] A. Calcaterra,[20a] G. F. Cao,[1,44] S. A. Cetin,[43b] J. Chai,[53c] J. F. Chang,[1,40] G. Chelkov,[24,b,c] G. Chen,[1] H. S. Chen,[1,44] J. C. Chen,[1] M. L. Chen,[1,40] P. L. Chen,[51] S. J. Chen,[30] X. R. Chen,[27] Y. B. Chen,[1,40] X. K. Chu,[32] G. Cibinetto,[21a] H. L. Dai,[1,40] J. P. Dai,[35,h] A. Dbeyssi,[14] D. Dedovich,[24] Z. Y. Deng,[1] A. Denig,[23] I. Denysenko,[24] M. Destefanis,[53a,53c] F. De Mori,[53a,53c] Y. Ding,[28] C. Dong,[31] J. Dong,[1,40] L. Y. Dong,[1,44] M. Y. Dong,[1,40,44] Z. L. Dou,[30] S. X. Du,[57] P. F. Duan,[1] J. Fang,[1,40] S. S. Fang,[1,44] Y. Fang,[1] R. Farinelli,[21a,21b] L. Fava,[53b,53c] S. Fegan,[23] F. Feldbauer,[23] G. Felici,[20a] C. Q. Feng,[50,40] E. Fioravanti,[21a] M. Fritsch,[23,14] C. D. Fu,[1] Q. Gao,[1] X. L. Gao,[50,40] Y. Gao,[42] Y. G. Gao,[6] Z. Gao,[50,40] I. Garzia,[21a] K. Goetzen,[10] L. Gong,[31] W. X. Gong,[1,40] W. Gradl,[23] M. Greco,[53a,53c] M. H. Gu,[1,40] Y. T. Gu,[12] A. Q. Guo,[1] R. P. Guo,[1,44] Y. P. Guo,[23] Z. Haddadi,[26] S. Han,[55] X. Q. Hao,[15] F. A. Harris,[45] K. L. He,[1,44] X. Q. He,[49] F. H. Heinsius,[4] T. Held,[4] Y. K. Heng,[1,40,44] T. Holtmann,[4] Z. L. Hou,[1] H. M. Hu,[1,44] T. Hu,[1,40,44] Y. Hu,[1] G. S. Huang,[50,40] J. S. Huang,[15] X. T. Huang,[34] X. Z. Huang,[30] Z. L. Huang,[28] T. Hussain,[52] W. Ikegami Andersson,[54] Q. Ji,[1] Q. P. Ji,[15] X. B. Ji,[1,44] X. L. Ji,[1,40] X. S. Jiang,[1,40,44] X. Y. Jiang,[31] J. B. Jiao,[34] Z. Jiao,[17] D. P. Jin,[1,40,44] S. Jin,[1,44] Y. Jin,[46] T. Johansson,[54] A. Julin,[47] N. Kalantar-Nayestanaki,[26] X. L. Kang,[1] X. S. Kang,[31] M. Kavatsyuk,[26] B. C. Ke,[5] T. Khan,[50,40] A. Khoukaz,[48] P. Kiese,[23] R. Kliemt,[10] L. Koch,[25] O. B. Kolcu,[43b,f] B. Kopf,[4] M. Kornicer,[45] M. Kuemmel,[4] M. Kuessner,[4] M. Kuhlmann,[4] A. Kupsc,[54] W. Kühn,[25] J. S. Lange,[25] M. Lara,[19] P. Larin,[14] L. Lavezzi,[53c] H. Leithoff,[23] C. Leng,[53c] C. Li,[54] Cheng Li,[50,40] D. M. Li,[57] F. Li,[1,40] F. Y. Li,[32] G. Li,[1] H. B. Li,[1,44] H. J. Li,[1,44] J. C. Li,[1] Jin Li,[33] K. J. Li,[41] Kang Li,[13] Ke Li,[34] Lei Li,[3] P. L. Li,[50,40] P. R. Li,[44,7] Q. Y. Li,[34] W. D. Li,[1,44] W. G. Li,[1] X. L. Li,[34] X. N. Li,[1,40] X. Q. Li,[31] Z. B. Li,[41] H. Liang,[50,40] Y. F. Liang,[37] Y. T. Liang,[25] G. R. Liao,[11] D. X. Lin,[14] B. Liu,[35,h] B. J. Liu,[1] C. X. Liu,[1] D. Liu,[50,40] F. H. Liu,[36] Fang Liu,[1] Feng Liu,[6] H. B. Liu,[12] H. M. Liu,[1,44] Huanhuan Liu,[1] Huihui Liu,[16] J. B. Liu,[50,40] J. P. Liu,[55] J. Y. Liu,[1,44] K. Liu,[42] K. Y. Liu,[28] Ke Liu,[6] L. D. Liu,[32] P. L. Liu,[1,40] Q. Liu,[44] S. B. Liu,[50,40] X. Liu,[27] Y. B. Liu,[31] Z. A. Liu,[1,40,44] Zhiqing Liu,[23] Y. F. Long,[32] X. C. Lou,[1,40,44] H. J. Lu,[17] J. G. Lu,[1,40] Y. Lu,[1] Y. P. Lu,[1,40] C. L. Luo,[29] M. X. Luo,[56] X. L. Luo,[1,40] X. R. Lyu,[44] F. C. Ma,[28] H. L. Ma,[1] L. L. Ma,[34] M. M. Ma,[1,44] Q. M. Ma,[1] T. Ma,[1] X. N. Ma,[31] X. Y. Ma,[1,40] Y. M. Ma,[34] F. E. Maas,[14] M. Maggiora,[53a,53c] Q. A. Malik,[52] Y. J. Mao,[32] Z. P. Mao,[1] S. Marcello,[53a,53c] Z. X. Meng,[46] J. G. Messchendorp,[26] G. Mezzadri,[21b] J. Min,[1,40] T. J. Min,[1] R. E. Mitchell,[19] X. H. Mo,[1,40,44] Y. J. Mo,[6] C. Morales Morales,[14] N. Yu. Muchnoi,[9,d] H. Muramatsu,[47] P. Musiol,[4] A. Mustafa,[4] Y. Nefedov,[24] F. Nerling,[10] I. B. Nikolaev,[9,d] Z. Ning,[1,40] S. Nisar,[8] S. L. Niu,[1,40] X. Y. Niu,[1,44] S. L. Olsen,[33,j] Q. Ouyang,[1,40,44] S. Pacetti,[20b] Y. Pan,[50,40] M. Papenbrock,[54] P. Patteri,[20a] M. Pelizaeus,[4] J. Pellegrino,[53a,53c] H. P. Peng,[50,40] K. Peters,[10,g] J. Pettersson,[54] J. L. Ping,[29] R. G. Ping,[1,44] A. Pitka,[23] R. Poling,[47] V. Prasad,[50,40] H. R. Qi,[2] M. Qi,[30] S. Qian,[1,40] C. F. Qiao,[44] N. Qin,[55] X. S. Qin,[4] Z. H. Qin,[1,40] J. F. Qiu,[1] K. H. Rashid,[52,i] C. F. Redmer,[23] M. Richter,[4] M. Ripka,[23] M. Rolo,[53c] G. Rong,[1,44] Ch. Rosner,[14] A. Sarantsev,[24,e] M. Savrié,[21b] C. Schnier,[4] K. Schoenning,[54] W. Shan,[32] M. Shao,[50,40] C. P. Shen,[2] P. X. Shen,[31] X. Y. Shen,[1,44] H. Y. Sheng,[1] J. J. Song,[34] W. M. Song,[34] X. Y. Song,[1] S. Sosio,[53a,53c] C. Sowa,[4] S. Spataro,[53a,53c] G. X. Sun,[1] J. F. Sun,[15] L. Sun,[55] S. S. Sun,[1,44] X. H. Sun,[1] Y. J. Sun,[50,40] Y. K. Sun,[50,40] Y. Z. Sun,[1] Z. J. Sun,[1,40] Z. T. Sun,[19] C. J. Tang,[37] G. Y. Tang,[1] X. Tang,[1] I. Tapan,[43c] M. Tiemens,[26] B. Tsednee,[22] I. Uman,[43d] G. S. Varner,[45] B. Wang,[1] B. L. Wang,[44] D. Wang,[32] D. Y. Wang,[32] Dan Wang,[44] K. Wang,[1,40] L. L. Wang,[1] L. S. Wang,[1] M. Wang,[34] Meng Wang,[1,44] P. Wang,[1] P. L. Wang,[1] W. P. Wang,[50,40] X. F. Wang,[42] Y. Wang,[38] Y. D. Wang,[14] Y. F. Wang,[1,40,44] Y. Q. Wang,[23] Z. Wang,[1,40] Z. G. Wang,[1,40] Z. H. Wang,[50,40,k] Z. Y. Wang,[1] Zongyuan Wang,[1,44] T. Weber,[23] D. H. Wei,[11] P. Weidenkaff,[23] S. P. Wen,[1] U. Wiedner,[4] M. Wolke,[54] L. H. Wu,[1] L. J. Wu,[1,44] Z. Wu,[1,40] L. Xia,[50,40] Y. Xia,[18] D. Xiao,[1] H. Xiao,[51] Y. J. Xiao,[1,44] Z. J. Xiao,[29] Y. G. Xie,[1,40] Y. H. Xie,[6] X. A. Xiong,[1,44] Q. L. Xiu,[1,40] G. F. Xu,[1] J. J. Xu,[1,44] L. Xu,[1] Q. J. Xu,[13] Q. N. Xu,[44] X. P. Xu,[38] L. Yan,[53a,53c] W. B. Yan,[50,40] W. C. Yan,[2] Y. H. Yan,[18] H. J. Yang,[35,h] H. X. Yang,[1] L. Yang,[55] Y. H. Yang,[30] Y. X. Yang,[11] M. Ye,[1,40] M. H. Ye,[7] J. H. Yin,[1] Z. Y. You,[41] B. X. Yu,[1,40,44] C. X. Yu,[31] J. S. Yu,[27] C. Z. Yuan,[1,44] Y. Yuan,[1] A. Yuncu,[43b,a] A. A. Zafar,[52] Y. Zeng,[18] Z. Zeng,[50,40] B. X. Zhang,[1] B. Y. Zhang,[1,40] C. C. Zhang,[1] D. H. Zhang,[1] H. H. Zhang,[41] H. Y. Zhang,[1,40] J. Zhang,[1,44] J. L. Zhang,[1] J. Q. Zhang,[1] J. W. Zhang,[1,40,44] J. Y. Zhang,[1] J. Z. Zhang,[1,44] K. Zhang,[1,44] L. Zhang,[42] S. Q. Zhang,[31] X. Y. Zhang,[34] Y. H. Zhang,[1,40] Y. T. Zhang,[50,40] Yang Zhang,[1] Yao Zhang,[1] Yu Zhang,[44] Z. H. Zhang,[6] Z. P. Zhang,[50] Z. Y. Zhang,[55] G. Zhao,[1] J. W. Zhao,[1,40] J. Y. Zhao,[1,44] J. Z. Zhao,[1,40] Lei Zhao,[50,40] Ling Zhao,[1] M. G. Zhao,[31] Q. Zhao,[1] S. J. Zhao,[57] T. C. Zhao,[1] Y. B. Zhao,[1,40] Z. G. Zhao,[50,40] A. Zhemchugov,[24,b] B. Zheng,[51] J. P. Zheng,[1,40] Y. H. Zheng,[44] B. Zhong,[29] L. Zhou,[1,40] X. Zhou,[55] X. K. Zhou,[50,40] X. R. Zhou,[50,40] X. Y. Zhou,[1] Y. X. Zhou,[12] J. Zhu,[31] J. Zhu,[41] K. Zhu,[1] K. J. Zhu,[1,40,44] S. Zhu,[1] S. H. Zhu,[49] X. L. Zhu,[42] Y. C. Zhu,[50,40] Y. S. Zhu,[1,44] Z. A. Zhu,[1,44] J. Zhuang,[1,40] B. S. Zou,[1] and J. H. Zou[1]

(BESIII Collaboration)

[1]*Institute of High Energy Physics, Beijing 100049, People's Republic of China*
[2]*Beihang University, Beijing 100191, People's Republic of China*
[3]*Beijing Institute of Petrochemical Technology, Beijing 102617, People's Republic of China*
[4]*Bochum Ruhr-University, D-44780 Bochum, Germany*
[5]*Carnegie Mellon University, Pittsburgh, Pennsylvania 15213, USA*
[6]*Central China Normal University, Wuhan 430079, People's Republic of China*
[7]*China Center of Advanced Science and Technology, Beijing 100190, People's Republic of China*
[8]*COMSATS Institute of Information Technology, Lahore, Defence Road, Off Raiwind Road, 54000 Lahore, Pakistan*
[9]*G. I. Budker Institute of Nuclear Physics SB RAS (BINP), Novosibirsk 630090, Russia*
[10]*GSI Helmholtzcentre for Heavy Ion Research GmbH, D-64291 Darmstadt, Germany*
[11]*Guangxi Normal University, Guilin 541004, People's Republic of China*
[12]*Guangxi University, Nanning 530004, People's Republic of China*
[13]*Hangzhou Normal University, Hangzhou 310036, People's Republic of China*
[14]*Helmholtz Institute Mainz, Johann-Joachim-Becher-Weg 45, D-55099 Mainz, Germany*
[15]*Henan Normal University, Xinxiang 453007, People's Republic of China*
[16]*Henan University of Science and Technology, Luoyang 471003, People's Republic of China*
[17]*Huangshan College, Huangshan 245000, People's Republic of China*
[18]*Hunan University, Changsha 410082, People's Republic of China*
[19]*Indiana University, Bloomington, Indiana 47405, USA*
[20a]*INFN Laboratori Nazionali di Frascati, I-00044, Frascati, Italy*
[20b]*INFN and University of Perugia, I-06100, Perugia, Italy*
[21a]*INFN Sezione di Ferrara, I-44122, Ferrara, Italy*
[21b]*University of Ferrara, I-44122, Ferrara, Italy*
[22]*Institute of Physics and Technology, Peace Ave. 54B, Ulaanbaatar 13330, Mongolia*
[23]*Johannes Gutenberg University of Mainz, Johann-Joachim-Becher-Weg 45, D-55099 Mainz, Germany*
[24]*Joint Institute for Nuclear Research, 141980 Dubna, Moscow region, Russia*
[25]*Justus-Liebig-Universitaet Giessen, II. Physikalisches Institut, Heinrich-Buff-Ring 16, D-35392 Giessen, Germany*
[26]*KVI-CART, University of Groningen, NL-9747 AA Groningen, The Netherlands*
[27]*Lanzhou University, Lanzhou 730000, People's Republic of China*
[28]*Liaoning University, Shenyang 110036, People's Republic of China*
[29]*Nanjing Normal University, Nanjing 210023, People's Republic of China*
[30]*Nanjing University, Nanjing 210093, People's Republic of China*
[31]*Nankai University, Tianjin 300071, People's Republic of China*
[32]*Peking University, Beijing 100871, People's Republic of China*
[33]*Seoul National University, Seoul, 151-747 Korea*
[34]*Shandong University, Jinan 250100, People's Republic of China*
[35]*Shanghai Jiao Tong University, Shanghai 200240, People's Republic of China*
[36]*Shanxi University, Taiyuan 030006, People's Republic of China*
[37]*Sichuan University, Chengdu 610064, People's Republic of China*
[38]*Soochow University, Suzhou 215006, People's Republic of China*
[39]*Southeast University, Nanjing 211100, People's Republic of China*
[40]*State Key Laboratory of Particle Detection and Electronics, Beijing 100049, Hefei 230026, People's Republic of China*
[41]*Sun Yat-Sen University, Guangzhou 510275, People's Republic of China*
[42]*Tsinghua University, Beijing 100084, People's Republic of China*
[43a]*Ankara University, 06100 Tandogan, Ankara, Turkey*
[43b]*Istanbul Bilgi University, 34060 Eyup, Istanbul, Turkey*
[43c]*Uludag University, 16059 Bursa, Turkey*
[43d]*Near East University, Nicosia, North Cyprus, Mersin 10, Turkey*
[44]*University of Chinese Academy of Sciences, Beijing 100049, People's Republic of China*
[45]*University of Hawaii, Honolulu, Hawaii 96822, USA*
[46]*University of Jinan, Jinan 250022, People's Republic of China*
[47]*University of Minnesota, Minneapolis, Minnesota 55455, USA*
[48]*University of Muenster, Wilhelm-Klemm-Str. 9, 48149 Muenster, Germany*
[49]*University of Science and Technology Liaoning, Anshan 114051, People's Republic of China*
[50]*University of Science and Technology of China, Hefei 230026, People's Republic of China*
[51]*University of South China, Hengyang 421001, People's Republic of China*
[52]*University of the Punjab, Lahore-54590, Pakistan*

[53a]University of Turin, I-10125, Turin, Italy
[53b]University of Eastern Piedmont, I-15121, Alessandria, Italy
[53c]INFN, I-10125, Turin, Italy
[54]Uppsala University, Box 516, SE-75120 Uppsala, Sweden
[55]Wuhan University, Wuhan 430072, People's Republic of China
[56]Zhejiang University, Hangzhou 310027, People's Republic of China
[57]Zhengzhou University, Zhengzhou 450001, People's Republic of China

Using $1.31\times10^9$ $J/\psi$ events collected by the BESIII detector at the BEPCII $e^+e^-$ collider, we report the first observation of the $h_1(1380)$ in $J/\psi\to\eta' h_1(1380)$ with a significance of more than ten standard deviations. The mass and width of the possible axial-vector strangeonium candidate $h_1(1380)$ are measured to be $M=(1423.2\pm2.1\pm7.3)$ MeV$/c^2$ and $\Gamma=(90.3\pm9.8\pm17.5)$ MeV. The product branching fractions, assuming no interference, are determined to be $\mathcal{B}(J/\psi\to\eta' h_1(1380))\times\mathcal{B}(h_1(1380)\to K^*(892)^+K^-+\text{c.c.})=(1.51\pm0.09\pm0.21)\times10^{-4}$ in $\eta' K^+K^-\pi^0$ mode and $\mathcal{B}(J/\psi\to\eta' h_1(1380))\times\mathcal{B}(h_1(1380)\to K^*(892)\bar{K}+\text{c.c.})=(2.16\pm0.12\pm0.29)\times10^{-4}$ in $\eta' K_S^0K^\pm\pi^\mp$ mode. The first uncertainties are statistical and the second are systematic. Isospin symmetry violation is observed in the decays $h_1(1380)\to K^*(892)^+K^-+\text{c.c.}$ and $h_1(1380)\to K^*(892)^0\bar{K}^0+\text{c.c.}$. Based on the measured $h_1(1380)$ mass, the mixing angle between the states $h_1(1170)$ and $h_1(1380)$ is determined to be $(35.9\pm2.6)°$, consistent with theoretical expectations.

## I. INTRODUCTION

The strangeonium spectrum is less well known at present compared to the charmonium and bottomonium spectra. Judging from its mass and large decay width to $K^*(892)\bar{K}+\text{c.c.}$ [1], the $h_1(1380)$ is a possible candidate for the $s\bar{s}$ partner of the $J^{PC}=1^{+-}$ axial-vector state $h_1(1170)$. Experimentally, the state $h_1(1380)$ has been observed by both the LASS [2] and Crystal Barrel [3] Collaborations, with masses and widths measured to be $M=(1380\pm20)$ MeV$/c^2$, $\Gamma=(80\pm30)$ MeV by LASS and $M=(1440\pm60)$ MeV$/c^2$, $\Gamma=(170\pm80)$ MeV by Crystal Barrel. Theoretically, the mass of the strangeonium $h_1(1380)$ is predicted to be $M=1468$ MeV$/c^2$ according to meson-mixing models [4,5], or $M=1386.42$ MeV$/c^2$, $(1415\pm13)$ MeV$/c^2$, 1470 MeV$/c^2$, $(1499\pm16)$ MeV$/c^2$ or 1511 MeV$/c^2$ according to quark models [6–10]. Assuming the $h_1(1380)$ is the $s\bar{s}$ partner of the $^1P_1$ state $h_1(1170)$, the $h_1(1380)$-$h_1(1170)$ mixing angle [11] can be determined from the masses of the $h_1(1380)$, $h_1(1170)$, $b_1(1235)$, $K_1(1400)$ and $K_1(1270)$, and the mixing angle between the $K_1(1400)$ and $K_1(1270)$ ($\theta_{K_1}$) [12]. Once the mixing angle is determined, it may shed light on the quark content of the $h_1(1380)$. In order to better understand the nature of the $h_1(1380)$, improved measurements are crucial.

With the huge charmonium data sets collected by the BESIII experiment, the strangeonium spectrum can be studied in charmonium decays. BESIII previously measured the mass and width of the $h_1(1380)$ as $M=(1412\pm9)$ MeV$/c^2$ and $\Gamma=(84\pm42)$ MeV via $\psi(3686)\to\gamma\chi_{cJ(J=1,2)}$, $\chi_{cJ(J=1,2)}\to\phi h_1(1380)$ and $h_1(1380)\to K^*(892)\bar{K}$, with $1.06\times10^8$ $\psi(3686)$ events collected at BESIII [13]. These results are consistent with those from the LASS and Crystal Barrel experiments [2,3], but are limited by the low statistics of the $\chi_{cJ}$ samples and large uncertainties from the interference of $h_1(1380)$ with the intermediate states $\phi(1680)$ and $\phi(1850)$. A more precise measurement would be useful for improving the understanding of the mass, quark content and corresponding mixing angle for the $h_1(1380)$.

[a]Also at Bogazici University, 34342 Istanbul, Turkey.
[b]Also at the Moscow Institute of Physics and Technology, Moscow 141700, Russia.
[c]Also at the Functional Electronics Laboratory, Tomsk State University, Tomsk, 634050, Russia.
[d]Also at the Novosibirsk State University, Novosibirsk, 630090, Russia.
[e]Also at the NRC "Kurchatov Institute", PNPI, 188300, Gatchina, Russia.
[f]Also at Istanbul Arel University, 34295 Istanbul, Turkey.
[g]Also at Goethe University Frankfurt, 60323 Frankfurt am Main, Germany.
[h]Also at Key Laboratory for Particle Physics, Astrophysics and Cosmology, Ministry of Education; Shanghai Key Laboratory for Particle Physics and Cosmology; Institute of Nuclear and Particle Physics, Shanghai 200240, People's Republic of China.
[i]Government College Women University, Sialkot—51310, Punjab, Pakistan.
[j]Present address: Center for Underground Physics, Institute for Basic Science, Daejeon 34126, Korea.
[k]Present address: HUAWEI TECHNOLOGIES CO., LTD., Shenzhen 518129, People's Republic of China.

In this paper, we present the first observation of $J/\psi \to \eta' h_1(1380)$, where $h_1(1380) \to K^*(892)\bar{K} + \text{c.c.} \to K^+K^-\pi^0/K_S^0 K^\pm \pi^\mp$, using a sample of $1.31 \times 10^9$ $J/\psi$ events [14,15].

## II. DETECTOR AND MONTE CARLO SIMULATION

The BESIII detector [16] is a magnetic spectrometer operating at BEPCII, a double-ring $e^+e^-$ collider with center of mass energies between 2.0 and 4.6 GeV. The cylindrical BESIII detector has an effective geometrical acceptance of 93% of $4\pi$. It is composed of a small cell helium-based main drift chamber (MDC) which provides momentum measurements for charged particles, a time-of-flight system (TOF) based on plastic scintillators that is used to identify charged particles, an electromagnetic calorimeter (EMC) made of CsI(Tl) crystals used to measure the energies of photons and electrons, and a muon system (MUC) made of resistive plate chambers (RPC). The momentum resolution of the charged particles is 0.5% at 1 GeV/$c$ in a 1 Tesla magnetic field. The energy loss ($dE/dx$) measurement provided by the MDC has a resolution of 6%, and the time resolution of the TOF is 80 ps (110 ps) in the barrel (end caps). The photon energy resolution is 2.5% (5%) at 1 GeV in the barrel (end caps) of the EMC.

A GEANT4 based [17] simulation software BOOST [18] is used to simulate the Monte Carlo (MC) samples. An inclusive $J/\psi$ MC sample is generated to estimate the backgrounds. The production of the $J/\psi$ resonance is simulated by the MC event generator KKMC [19], while the decays are generated by BESEVTGEN [20] for known decays modes with branching fractions according to the world average values [1], and by the LUNDCHARM model [21] for the remaining unknown decays. Exclusive MC samples are generated to determine the detection efficiencies of the signal processes and optimize event selection criteria.

## III. EVENT SELECTION

For $J/\psi \to \eta' K^+K^-\pi^0$ with $\eta' \to \pi^+\pi^-\eta$, $\eta \to \gamma\gamma$ and $\pi^0 \to \gamma\gamma$, candidate events are required to have four charged tracks with zero net charge and at least four photons. Each charged track is required to be within the polar angle range $|\cos\theta| < 0.93$ and must pass within 10 cm (1 cm) of the interaction point in the beam (radial) direction. Information from TOF and $dE/dx$ measurements is combined to form particle identification (PID) confidence levels for the $\pi$, $K$, and $p$ hypotheses. Each track is assigned the particle type corresponding to the hypothesis with the highest confidence level. Two oppositely charged kaons and pions are required for each event. Photon candidates are reconstructed from isolated clusters of energy deposits in the EMC and must have an energy of at least 25 MeV for barrel showers ($|\cos\theta| < 0.8$), or 50 MeV for end cap showers ($0.86 < |\cos\theta| < 0.92$). The energy deposited in nearby TOF counters is also included. EMC cluster timing requirements ($0 \leq t \leq 14$ in units of 50 ns) are used to suppress electronics noise and energy deposits unrelated to the event.

To improve the momentum and energy resolution and suppress background events, a four-constraint (4C) kinematic fit imposing energy-momentum conservation is performed under the hypothesis $J/\psi \to \gamma\gamma\gamma\gamma\pi^+\pi^-K^+K^-$, and a requirement of $\chi^2_{4C} < 100$ is imposed. For events with more than four photon candidates, the combination with the smallest $\chi^2_{4C}$ is retained.

Photon pairs corresponding to the best $\pi^0\eta$, $\pi^0\pi^0$ and $\eta\eta$ candidates are selected using the quantities $\chi^2_{\alpha\beta} = (M_{\gamma_1\gamma_2} - m_\alpha)^2/\sigma^2_\alpha + (M_{\gamma_3\gamma_4} - m_\beta)^2/\sigma^2_\beta$, where $\alpha\beta = \pi^0\eta, \pi^0\pi^0$, or $\eta\eta$ and each mass resolution $\sigma_{\alpha(\beta)}$ is obtained from the MC simulation. Only the combination with $\chi^2_{\pi^0\eta} < \chi^2_{\pi^0\pi^0}$ and $\chi^2_{\pi^0\eta} < \chi^2_{\eta\eta}$ is retained. The $\pi^0$ and $\eta$ candidates are selected by requiring $|M(\gamma\gamma) - m_{\pi^0}| < 0.02$ GeV/$c^2$ and $|M(\gamma\gamma) - m_\eta| < 0.03$ GeV/$c^2$, respectively. The $\pi^+\pi^-\eta$ invariant mass distribution for the selected events is shown in Fig. 1, where an $\eta'$ peak is evident. The peak around 1.3 GeV/$c^2$ is due to $f_1(1285)$ or $\eta(1295)$ decays. Events with $|M(\pi^+\pi^-\eta) - m_{\eta'}| < 0.03$ GeV/$c^2$ are selected for further analysis. Here, $m_{\pi^0}$, $m_\eta$, and $m_{\eta'}$ are the nominal masses of $\pi^0$, $\eta$, and $\eta'$ [1].

After the above selection criteria, the distribution of the invariant mass of $K^+\pi^0$ versus that of $K^-\pi^0$ found in the data is shown in Fig. 2(a). Bands for the $K^*(892)^\pm$ are evident, indicating that the $J/\psi \to \eta' K^*(892)^+K^- + \text{c.c.}$ process is dominant. Figures 2(b) and 2(c) show the projections of the $K^+\pi^0$ and $K^-\pi^0$ invariant masses, respectively.

Potential background processes to $J/\psi \to \eta' K^*(892)^+K^- + \text{c.c.}$ are studied using an inclusive sample of $1.2 \times 10^9$ $J/\psi$ events. Simulated events are subject to the same selection procedure as that applied to the data.

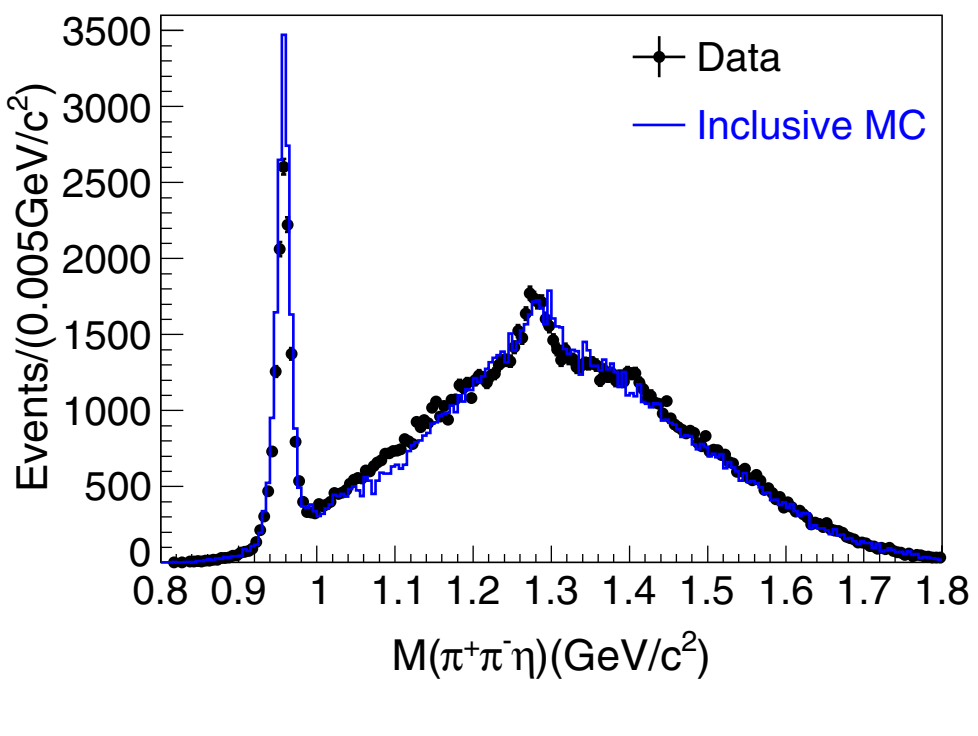

FIG. 1. Distribution of the $\pi^+\pi^-\eta$ invariant mass in the $\eta' K^+K^-\pi^0$ mode. The dots with error bars are data and the histogram is the inclusive MC sample.

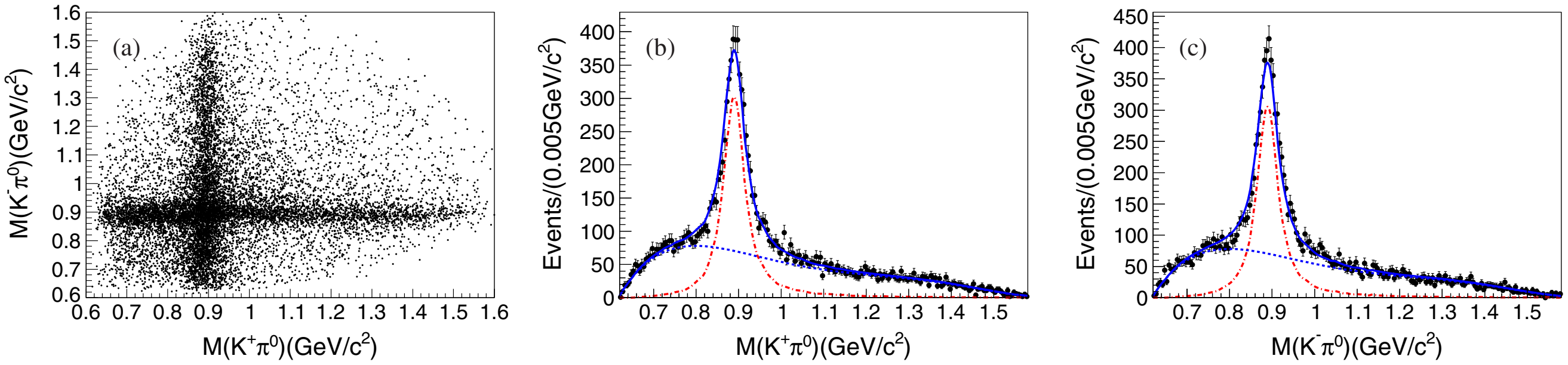

FIG. 2. (a) Scatter plot of the $K^+\pi^0$ invariant mass versus that of $K^-\pi^0$ in selected data events. Fits to the (b) $M(K^+\pi^0)$ and (c) $M(K^-\pi^0)$ distributions, where the dots with error bars are data, the solid curves are the total fit results, the dashed curves indicate backgrounds and the dotted-dashed curves are $K^*(892)$ signal shapes.

No significant peaking background sources are identified. The dominant backgrounds stem from $J/\psi \to \phi\eta\eta \to K^+K^-\eta\pi^+\pi^-\pi^0$ and $J/\psi \to \phi f_0(1710)$ where $\phi \to K^+K^-$ and $f_0(1710) \to \eta\eta \to \eta\pi^+\pi^-\pi^0$. The possible peaking backgrounds are considered in the $\eta'$ sideband regions defined as $0.035 < |M(\pi^+\pi^-\eta) - m_{\eta'}| < 0.065\ \mathrm{GeV}/c^2$. The peaking contribution in the $K^*(892)^\pm$ signal region is found to be small and will be taken into account in the systematic uncertainties.

For $J/\psi \to \eta' K_S^0 K^\pm\pi^\mp$ with $\eta' \to \pi^+\pi^-\eta$, $\eta \to \gamma\gamma$ and $K_S^0 \to \pi^+\pi^-$, candidate events are required to have six charged tracks with zero net charge and at least two photons. Each charged track and photon candidate is reconstructed as described above except for the $\pi^+\pi^-$ pair from $K_S^0$. The $K_S^0$ candidates are reconstructed from all combinations of pairs of oppositely charged tracks, assuming each of the two tracks is a pion. A secondary vertex fit is performed and the fit $\chi^2$ is required to be less than 100. If more than one $K_S^0$ candidate is reconstructed in an event, the one with the minimum $|M(\pi^+\pi^-) - m_{K_S^0}|$ is selected for further analysis. The $K_S^0$ candidates are further required to satisfy $|M(\pi^+\pi^-) - m_{K_S^0}| < 0.01\ \mathrm{GeV}/c^2$. Here, $m_{K_S^0}$ is the nominal mass of $K_S^0$ [1]. The other four charged tracks must be identified as three pions and one kaon according to PID information.

For each event, a 4C kinematic fit is performed under the hypothesis of $J/\psi \to \gamma\gamma\pi^+\pi^-K_S^0K^\pm\pi^\mp$, where the $K_S^0$ candidate is included with the parameters obtained from the second vertex fit. A requirement of $\chi^2_{4C} < 100$ is imposed. The $\eta$ candidate is selected by requiring $|M(\gamma\gamma) - m_\eta| < 0.03\ \mathrm{GeV}/c^2$. The $\pi^+\pi^-\eta$ mass distribution is shown in Fig. 3, choosing the oppositely charged pion combination which gives the $\pi^+\pi^-\eta$ mass closest to the nominal $\eta'$ mass. The $\eta'$ signal is observed and selected with the requirement of $|M(\pi^+\pi^-\eta) - m_{\eta'}| < 0.03\ \mathrm{GeV}/c^2$. Similarly to that of Fig. 1, the peak around 1.3 GeV/$c^2$ is due to $f_1(1285)$ or $\eta(1295)$ decays.

After the above selection criteria, the distribution of the invariant mass of $K_S^0\pi^\pm$ versus that of $K^\pm\pi^\mp$ found in data is shown in Fig. 4(a). Bands for the $K^*(892)^\pm$ and $K^*(892)^0$ ($\bar{K}^*(892)^0$) are evident, indicating that the $J/\psi \to \eta' K^*(892)\bar{K}$ + c.c. process is dominant. Figures 4(b) and 4(c) show the projections of the $K_S^0\pi^\pm$ and $K^\pm\pi^\mp$ invariant masses, respectively.

Similarly to that of $J/\psi \to \eta' K^*(892)^+K^-$ + c.c., potential background processes to $J/\psi \to \eta' K^*(892)\bar{K}$ + c.c. are studied using an inclusive sample of $1.2 \times 10^9$ $J/\psi$ events. No significant peaking background sources are identified. The dominant backgrounds stem from the four-body decay of $J/\psi \to \eta' K_S^0 K^\pm\pi^\mp$. The possible peaking backgrounds are considered in the $\eta'$ sideband region defined as $0.035 < |M(\pi^+\pi^-\eta) - m_{\eta'}| < 0.065\ \mathrm{GeV}/c^2$. The peaking contribution in the $K^*(892)^\pm$ and $K^*(892)^0$ ($\bar{K}^*(892)^0$) signal regions is found to be small and will be taken into account in the systematic uncertainties.

## IV. EXTRACTION OF BRANCHING FRACTIONS

To determine the signal yields of $J/\psi \to \eta' K^*(892)\bar{K}$+ c.c., a simultaneous unbinned maximum likelihood fit is performed to the $M(K^+\pi^0)$ and $M(K^-\pi^0)$ spectra for the $K^+K^-\pi^0$ mode. The signal shapes are taken directly from the corresponding MC simulation, where an interpolation is applied to extract a smoothed shape. The backgrounds are described with fifth-order Chebychev polynomial functions. In the $K^+K^-\pi^0$ mode, the efficiencies of the charge conjugated channels are found to be consistent within the statistical uncertainties, and the number of signal events containing a $K^*(892)^+$ or a $K^*(892)^-$ is constrained to be the same in the fit. The fit yields a total of $5066 \pm 79$ events, as shown in Figs. 2(b) and 2(c). The goodness of the fits are found to be $\chi^2/\mathrm{ndf} = 172/186 = 0.92$ in $M(K^+\pi^0)$ spectrum and $189/186 = 1.02$ in $M(K^-\pi^0)$ spectrum, where the ndf is the number of degrees of freedom. In the $K_S^0K^\pm\pi^\mp$ mode, a similar simultaneous fit is performed to the $M(K_S^0\pi^\pm)$ and $M(K^\pm\pi^\mp)$ spectra. The fit yields $7749 \pm 134$ $K^*(892)^\pm$ and $8268 \pm 137$ $K^*(892)^0$ or $\bar{K}^*(892)^0$ events, as shown in Figs. 4(b) and 4(c). The goodness of the fits are $\chi^2/\mathrm{ndf} = 211/181 = 1.17$ in $M(K_S^0\pi^\pm)$ spectrum and $251/181 = 1.39$ in $M(K^\pm\pi^\mp)$ spectrum. Here, the uncertainties are statistical only.

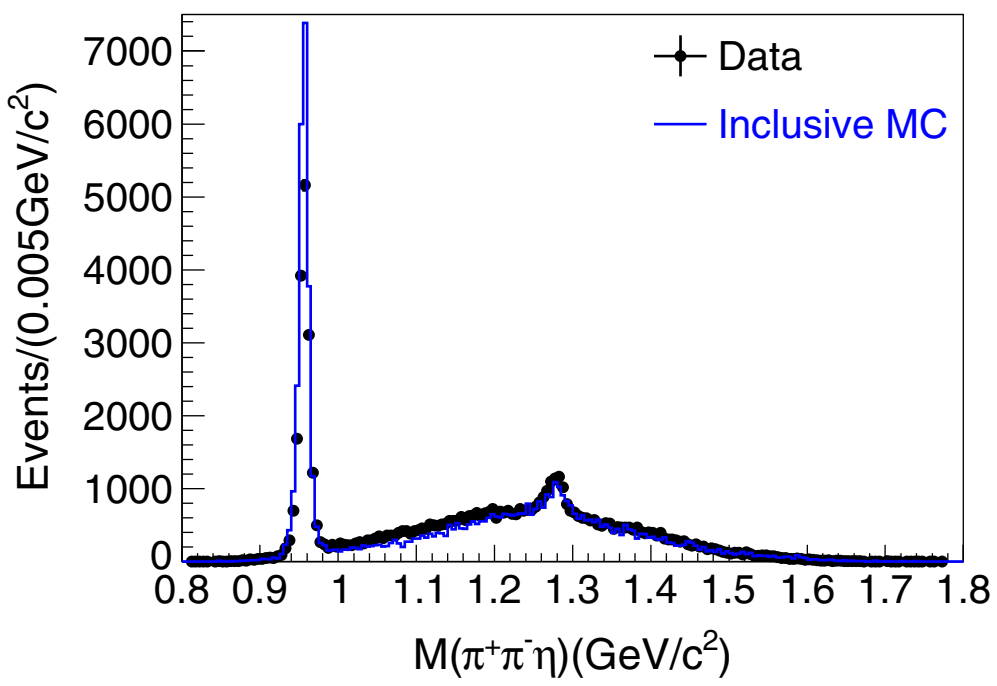

FIG. 3. Distribution of the $\pi^+\pi^-\eta$ invariant mass closest to the $\eta'$ mass in the $\eta' K_S^0 K^\pm\pi^\mp$ mode. The dots with error bars are data and the histogram is the inclusive MC sample.

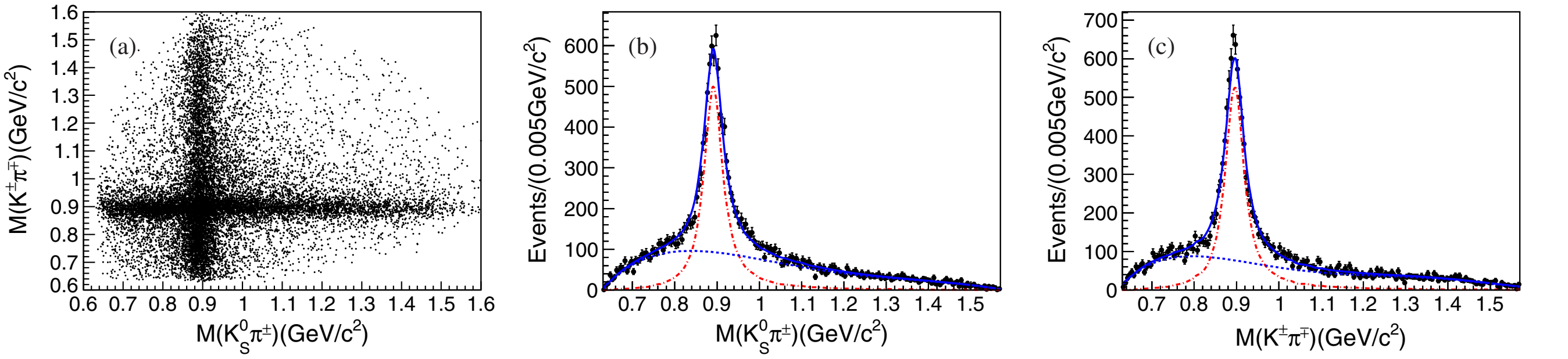

FIG. 4. (a) Scatter plot of the $K_S^0\pi^\pm$ invariant mass versus that of $K^\pm\pi^\mp$. Fits to the (b) $M(K_S^0\pi^\pm)$ and (c) $M(K^\pm\pi^\mp)$ distributions, where the dots with error bars are data, the solid curves are the total fit results, the dashed curves indicate background and the dotted-dashed curves are $K^*(892)$ signal shapes.

The branching fractions are calculated with $\mathcal{B}(J/\psi \to \eta' K^*\bar{K} + \mathrm{c.c.}) = N^{\mathrm{obs}}/(N_{J/\psi} \times \mathcal{B} \times \epsilon)$, where $N^{\mathrm{obs}}$ is the total number of signal events; $N_{J/\psi}$ is the number of $J/\psi$ decays [14,15]; $\epsilon$ is the selection efficiency obtained from a phase space MC simulation; and $\mathcal{B}$ is the product of branching fractions of intermediate states. Considering the negligible differences for the final states with and without the $h_1(1380)$, the signal efficiencies are obtained using exclusive MC samples without the $h_1(1380)$. The selection efficiencies are 9.3% and 10.3% (9.8%) for the decay modes $\eta' K^+K^-\pi^0$ and $\eta' K_S^0K^\pm\pi^\mp$ with an intermediate $K^*(892)^\pm$ ($K^*(892)^0/\bar{K}^*(892)^0$), respectively. The measured branching fractions are $\mathcal{B}(J/\psi \to \eta' K^*(892)^+K^- + \mathrm{c.c.}) = (1.50 \pm 0.02) \times 10^{-3}$ for the $\eta' K^+K^-\pi^0$ mode and $\mathcal{B}(J/\psi \to \eta' K^*(892)^+K^- + \mathrm{c.c.}) = (1.47 \pm 0.03) \times 10^{-3}$, $\mathcal{B}(J/\psi \to \eta' K^*(892)^0\bar{K}^0 + \mathrm{c.c.}) = (1.66 \pm 0.03) \times 10^{-3}$ for the $\eta' K_S^0K^\pm\pi^\mp$ mode. Here, the uncertainties are statistical only.

## V. STUDY OF INTERMEDIATE STATES

Intermediate states are studied by examining the $K\bar{K}\pi$ invariant mass distributions. The $K^*(892)$ signals are selected using $|M(K^\pm\pi^0) - m_{K^*(892)^\pm}| < 0.15$ GeV/$c^2$ in the $\eta' K^+K^-\pi^0$ mode and $|M(K_S^0\pi^\pm) - m_{K^*(892)^\pm}| < 0.15$ GeV/$c^2$ or $|M(K^\pm\pi^\mp) - m_{K^*(892)^0}| < 0.15$ GeV/$c^2$ in the $\eta' K_S^0K^\pm\pi^\mp$ mode. Here, $m_{K^*(892)^\pm}$ and $m_{K^*(892)^0}$ are the nominal masses of $K^*(892)^\pm$ and $K^*(892)^0$ [1].

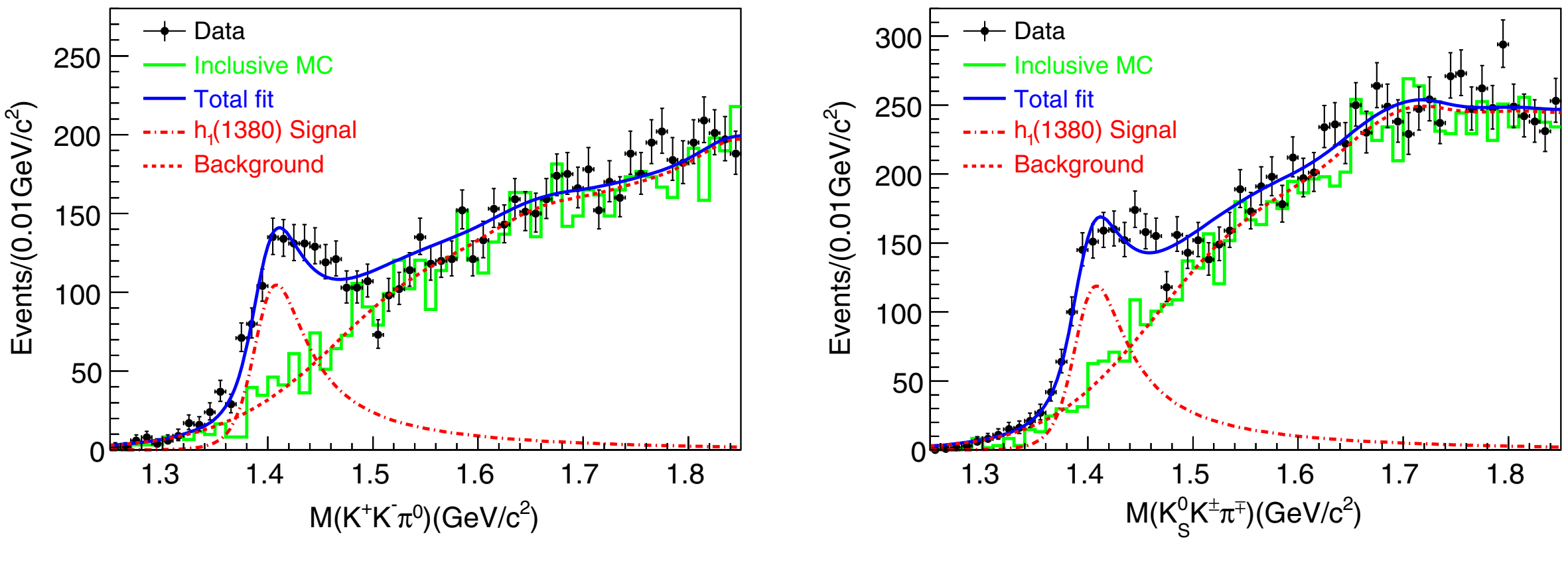

FIG. 5. Fits to the $M(K^*(892)\bar{K})$ distributions as described in the text. The dots with error bars are data; the solid curves show the total fits; the dotted-dashed curves are $h_1(1380)$ signals; and the solid histograms are from inclusive MC samples with $h_1(1380)$ signals removed; the short-dashed curves are the backgrounds with its shape modeled from the solid histograms.

Figure 5 shows the selected $K^+K^-\pi^0$ and $K_S^0K^\pm\pi^\mp$ invariant mass distributions after the $K^*(892)$ selection, where a distinct peak near the $K^*(892)\bar{K}$ mass threshold is observed. Potential background processes are studied with inclusive $J/\psi$ MC samples and sideband events from $\pi^0$, $\eta'$ and $K^*(892)$ ($\bar{K}^*(892)$). None of the background sources produces an enhancement at the $K^*(892)\bar{K}$ mass threshold region. The dominant backgrounds are from three body decays of $J/\psi \to \eta' K^*(892)\bar{K} + \text{c.c.}$. Assuming that this threshold enhancement comes from an intermediate state and taking into account its mass, its decays through $K^*(892)\bar{K}$, and charge parity conservation, the most likely assignment for this structure is the $h_1(1380)$ ($J^{PC} = 1^{+-}$) [1].

To characterize the observed enhancement and determine the signal yields, a simultaneous unbinned maximum likelihood fit is performed to the $M(K^*(892)\bar{K})$ distributions in the $K^+K^-\pi^0$ and $K_S^0K^\pm\pi^\mp$ modes with a common mass and width for the $h_1(1380)$ signal. The signal shape is parameterized using a relativistic $S$-wave Breit-Wigner function with a mass-dependent width multiplied by a phase space factor $q$,

$$\left|\frac{\sqrt{m\Gamma(m)}}{m^2 - m_0^2 + im\Gamma(m)}\right|^2 \times q \quad (1)$$

where $\Gamma(m) = \Gamma_0(\frac{m_0}{m})(\frac{p}{p_0})^{2l+1}$, $l = 0$ is the orbital momentum, $m$ is the reconstructed mass of $K^*(892)\bar{K}$, $m_0$ and $\Gamma_0$ are the nominal resonance mass and width, $q$ is the $\eta'$ momentum in the $J/\psi$ rest frame, $p$ is the $\bar{K}$ momentum in the rest frame of the $K^*(892)\bar{K}$ system, and $p_0$ is the $\bar{K}$ momentum in the resonance rest frame at $m = m_0$. The large total decay widths of the $K^*(892)$ are taken into account by convolving the momentum of the $\bar{K}$ with the invariant mass distribution of the $K^*(892)$ [22]. The mass resolution, fixed to the MC simulated value of 6.0 MeV$/c^2$, is taken into account by convolving the signal shape with a Gaussian function. In the fit, the background shape is modeled from inclusive MC based on kernel estimation [23] and its magnitude is allowed to vary. The possible interference between the signal and background is neglected in the fit.

The fit yields a mass of $(1423.2 \pm 2.1)$ MeV$/c^2$ and a width of $(90.3 \pm 9.8)$ MeV, as shown in Fig. 5. The fit qualities ($\chi^2/\text{ndf}$, with ndf $= 56$) are 1.36 for the $K^+K^-\pi^0$ mode and 1.05 for the $K_S^0K^\pm\pi^\mp$ mode. The numbers of the fitted $h_1(1380)$ signal events are $1054 \pm 60$ and $1195 \pm 68$ for the $K^+K^-\pi^0$ and $K_S^0K^\pm\pi^\mp$ modes, respectively. The product branching fractions are $\mathcal{B}(J/\psi \to \eta' h_1(1380)) \times \mathcal{B}(h_1(1380) \to K^*(892)^+K^- + \text{c.c.}) = (1.51 \pm 0.09) \times 10^{-4}$ in the $\eta' K^+K^-\pi^0$ mode and $\mathcal{B}(J/\psi \to \eta' h_1(1380)) \times \mathcal{B}(h_1(1380) \to K^*(892)\bar{K} + \text{c.c.}) = (2.16 \pm 0.12) \times 10^{-4}$ in the $\eta' K_S^0K^\pm\pi^\mp$ mode. Here, the uncertainties are statistical only. The statistical significance is calculated by comparing the fit likelihoods with and without the $h_1(1380)$ signal with the change on the number of degrees of freedom considered. The differences due to the fit uncertainties by changing the fit range, the signal shape, or the background shape are included into the systematic uncertainties. In all cases, the significance is found to be larger than $10\sigma$. According to isospin symmetry, $\mathcal{B}(h_1(1380) \to K^*(892)^+K^- + \text{c.c.})$ should be equal to $\mathcal{B}(h_1(1380) \to K^*(892)^0\bar{K}^0 + \text{c.c.})$. However, considering the mass differences between the charged and neutral $K$ and $K^*(892)$ mesons ($\Delta m_K = 3.97$ MeV$/c^2$, and $\Delta m_{K^*(892)} = 4.15$ MeV$/c^2$ [1]) and the fact that the $h_1(1380)$ state resides near the $K^*(892)\bar{K}$ threshold, isospin symmetry breaking effects are expected [24,25].

We also fit the $K^*(892)\bar{K}$ invariant mass distribution allowing interference between the $h_1(1380)$ signal and the nonresonant background. The amplitude of nonresonant background is extracted by a fit to the inclusive MC with the sixth-order of Chebychev polynomial function. The magnitude of the background probability density function and phase angle is allowed to vary, and the lowest negative likelihood corresponds to constructive interference. The final fit and the individual contribution of each component are shown in Fig. 6. The fitted mass and width of the $h_1(1380)$ are $M = (1441.7 \pm 4.9)\ \mathrm{MeV}/c^2$ and $\Gamma = (111.5 \pm 12.8)$ MeV. In this analysis, the fit results without considering interference are taken as the nominal values.

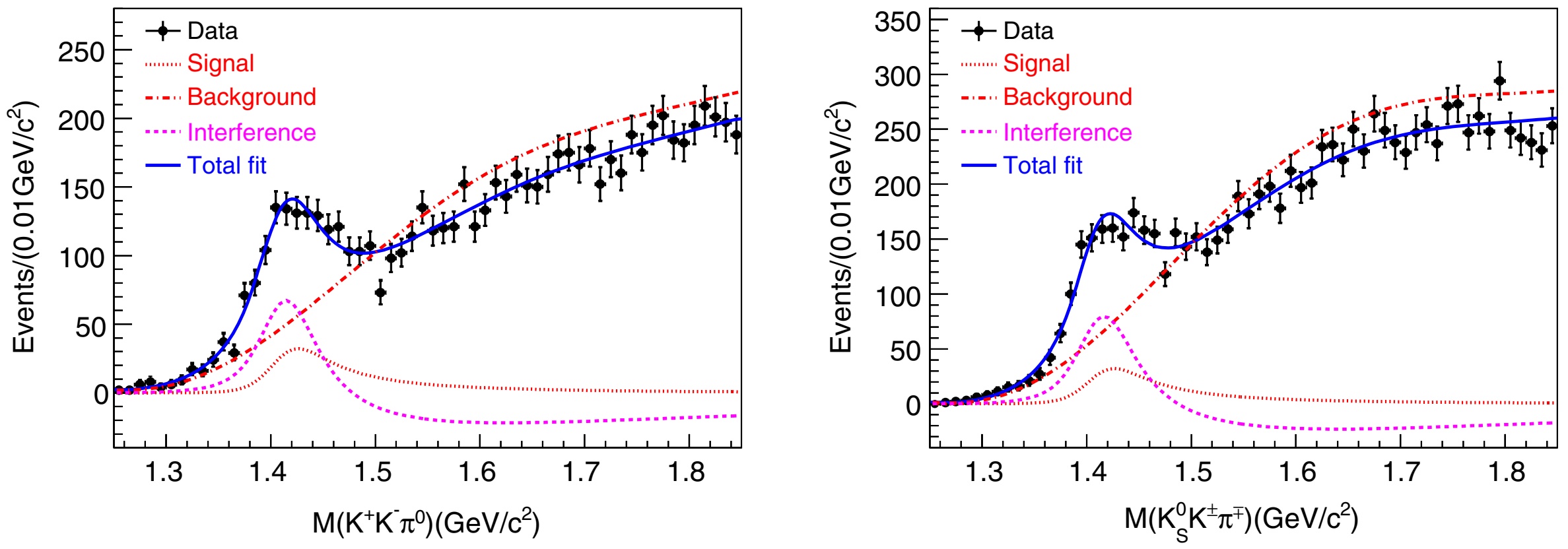

FIG. 6. Fits to the $M(K^*(892)\bar{K})$ distributions with interference between signal and background. Dots with error bars are data; the solid curves show the total fits; the dot-dashed curves are the background; the dotted curves are the $h_1(1380)$ signal; and the short-dashed curves are the interference between signal and background.

## VI. SYSTEMATIC UNCERTAINTIES

Sources of systematic uncertainties for the $h_1(1380)$ resonance parameters include the mass calibration, parameterizations of the signal and background shapes, fit range and mass resolution. The uncertainty from the mass calibration is estimated using the difference between the measured $\eta'$ mass $(956.82 \pm 0.11)\ \mathrm{MeV}/c^2$ and the nominal value $(957.78 \pm 0.06)\ \mathrm{MeV}/c^2$ [1]. The uncertainty due to the mass resolution is estimated by varying the resolution from 6.0 MeV to 6.7 MeV, as a 11% difference is seen between data and simulation for the $\eta'$ mass resolution. For the systematic uncertainty associated with the signal shape, an alternative fit is performed by assuming a $P$-wave between the $\eta'$ and the $h_1(1380)$. The uncertainty due to the background shape is determined by changing the inclusive MC shape to a third-order Chebychev polynomial function. The fit range is varied to determine the associated uncertainty. Finally the individual uncertainties are summarized in Table I. Assuming all sources of systematic uncertainty are independent and adding them in quadrature, the total systematic uncertainty is 7.3 $\mathrm{MeV}/c^2$ for the mass, and 17.5 MeV for the width of the $h_1(1380)$.

Systematic uncertainties in the branching fraction measurements come from the uncertainties in the number of $J/\psi$ events, tracking efficiency, particle identification, photon detection, $K_S^0$ reconstruction, kinematic fit, mass window requirements, fitting procedure, peaking background estimation, and the branching fractions of intermediate state decays.

In Refs. [14,15], the number of $J/\psi$ events is determined with an uncertainty of 0.6%. The uncertainty of the tracking efficiency is estimated to be 1.0% for each pion and kaon from a study of the control samples $J/\psi \to K_S^0 K^\pm \pi^\mp$ and $K_S^0 \to \pi^+\pi^-$ [26]. With the control samples, the uncertainty from PID is estimated to be 2.0% for each charged pion and kaon. The uncertainty due to photon detection is 1.0% per photon, as obtained from a study of the high-purity control sample of $J/\psi \to \rho\pi$ [27]. For $K_S^0$ reconstruction, the uncertainty is studied with a control sample of $J/\psi \to K^*(892)^\pm K^\mp \to K_S^0 K^\pm \pi^\mp$. A conservative value of 3.5% is taken as the systematic uncertainty. The uncertainty associated with the kinematic fit comes from the inconsistency between data and MC simulation of the track helix parameters and the error matrices. Following the procedure described in Ref. [28], we take the difference between the efficiencies with and without the helix parameter correction as the systematic uncertainty, which is 2.8% in the $\eta' K^+ K^- \pi^0$ mode and 1.6% in the $\eta' K_S^0 K^\pm \pi^\mp$ mode. The uncertainties arising from the $\pi^0$, $\eta$, $\eta'$ and $K_S^0$ selection are estimated by varying the mass window requirements. To estimate the uncertainties from the choice of signal shape, background shape and fit range, for the $K^*(892)$ signal fit, the signal shape is changed from the MC shape to a Breit-Wigner function convolved with a Gaussian function; the background shape is varied from a polynomial function to the MC shape plus the non-$\eta'$ sideband, and the fit range is also varied; for the $h_1(1380)$ signal fit, the methods are following the $h_1(1380)$ resonance parameters study described above. The peaking background from the $K^*(892)$ is estimated using the non-$\eta'$ sidebands. The uncertainties associated with the branching fractions of intermediate states are taken from the Particle Data Group [1]. The total systematic uncertainties in the branching fractions are determined to be 14.1% and 12.4% for $\mathcal{B}(J/\psi \to \eta' h_1(1380)) \times \mathcal{B}(h_1(1380) \to K^*(892)^+ K^- + \mathrm{c.c.})$ and $\mathcal{B}(J/\psi \to \eta' K^*(892)^+ K^- + \mathrm{c.c.})$, respectively, for the $\eta' K^+ K^- \pi^0$ final states, and 13.3%, 11.8% and 12.9% for $\mathcal{B}(J/\psi \to \eta' h_1(1380)) \times \mathcal{B}(h_1(1380) \to K^*(892)\bar{K} + \mathrm{c.c.})$, $\mathcal{B}(J/\psi \to \eta' K^*(892)^+ K^- + \mathrm{c.c.})$ and $\mathcal{B}(J/\psi \to \eta' K^* \times (892)^0 \bar{K}^0 + \mathrm{c.c.})$, respectively, for $\eta' K_S^0 K^\pm \pi^\mp$ final states, as summarized in Table II.

TABLE I. Systematic uncertainties for the $h_1(1380)$ resonance parameters.

| Source | $M$ (MeV/$c^2$) | $\Gamma$ (MeV) |
|---|---|---|
| Mass calibration | 1.1 | ... |
| Mass resolution | ... | 1.8 |
| Signal shape | 2.1 | 4.7 |
| Background shape | 6.8 | 16.7 |
| Fit range | 1.1 | 1.4 |
| Total | 7.3 | 17.5 |

## VII. MIXING ANGLE BETWEEN $h_1(1170)$ AND $h_1(1380)$

The mixing angle $\theta_{^1p_1}$ between the $h_1(1170)$ and $h_1(1380)$ is calculated with the relation [11]

$$\tan\theta_{^1p_1} = \frac{m^2_{^1p_1} - m^2_{h'_1}}{\sqrt{m^2_{^1p_1}(m^2_{h_1} + m^2_{h'_1} - m^2_{^1p_1}) - m^2_{h_1} m^2_{h'_1}}}, \quad (2)$$

where $m'_{h_1}$ and $m_{h_1}$ are the masses of $h_1(1380)$ and $h_1(1170)$, respectively, and $m^2_{^1p_1}$ is the mass squared of the octet state $^1p_1$, applying the Gell-Mann-Okubo relations [29], obtained as

$$m_8^2(^1p_1) \equiv m^2_{^1p_1} = \frac{1}{3}(4m^2_{K_{1B}} - m^2_{b1}). \quad (3)$$

TABLE II. Systematic uncertainties in the branching fractions of $\mathcal{B}(J/\psi \to \eta' h_1(1380)) \times \mathcal{B}(h_1(1380) \to K^*(892)\bar{K} + \text{c.c.})$ and $\mathcal{B}(J/\psi \to \eta' K^*(892)\bar{K} + \text{c.c.})$ (in %).

| Source | $\eta' h_1(1380)$ $(K^+K^-\pi^0)$ | $\eta' h_1(1380)$ $(K^0_S K^\pm \pi^\mp)$ | $\eta' K^{*\pm}K^\mp$ $(K^+K^-\pi^0)$ | $\eta' K^{*\pm}K^\mp$ $(K^0_S K^\pm \pi^\mp)$ | $\eta' K^{*0}\bar{K}^0 + \eta' \bar{K}^{*0}K^0$ $(K^0_S K^\pm \pi^\mp)$ |
|---|---|---|---|---|---|
| Number of $J/\psi$ | 0.6 | 0.6 | 0.6 | 0.6 | 0.6 |
| MDC tracking | 4.0 | 6.0 | 4.0 | 6.0 | 6.0 |
| Photon detection | 4.0 | 2.0 | 4.0 | 2.0 | 2.0 |
| Particle identification | 8.0 | 8.0 | 8.0 | 8.0 | 8.0 |
| $K^0_S$ reconstruction | – | 3.5 | ⋯ | 3.5 | 3.5 |
| 4C kinematic fit | 2.8 | 1.6 | 2.8 | 1.6 | 1.6 |
| $\pi^0$ selection | 2.2 | ⋯ | 0.3 | ⋯ | ⋯ |
| $\eta$ selection | 1.3 | 0.4 | 0.2 | 0.1 | 0.1 |
| $\eta'$ selection | 3.4 | 3.6 | 0.8 | 0.6 | 0.5 |
| $K^0_S$ selection | ⋯ | 0.6 | ⋯ | 0.9 | 0.2 |
| $K^*(892)$ selection | 0.3 | 0.4 | ⋯ | ⋯ | ⋯ |
| Signal shape | 5.3 | 5.5 | 5.3 | 3.2 | 4.1 |
| Background shape | 6.0 | 2.6 | 3.8 | 1.6 | 5.0 |
| Fit range | 3.0 | 2.2 | 0.8 | 0.4 | 0.4 |
| $K^*(892)$ peaking background | ⋯ | ⋯ | 1.4 | 1.7 | 1.6 |
| Branching fraction | 1.7 | 1.7 | 1.7 | 1.7 | 1.7 |
| Total | 14.1 | 13.3 | 12.4 | 11.8 | 12.9 |

Finally, $m_{K_{1B}}$ is the mass of the flavor eigenstate $K_{1B}$ as obtained from the relation

$$m^2_{K_{1B}} = m^2_{K_1(1400)}\sin^2\theta_{K_1} + m^2_{K_1(1270)}\cos^2\theta_{K_1}. \quad (4)$$

Based on the $h_1(1380)$ mass measured in this analysis and the masses of the $h_1(1170)$, $b_1(1235)$, $K_1(1400)$ and $K_1(1270)$ taken from the Particle Data Group [1], and the $K_{1A} - K_{1B}$ mixing angle, $\theta_{K_1} = 34°$ [11], the mixing angle between the $h_1(1170)$ and $h_1(1380)$ is determined to be $\theta_{^1p_1} = (35.9 \pm 2.6)°$, assuming $h_1(1380)$ is a prime $s\bar{s}$ state [11] and considering it decays to $K^*(892)\bar{K}$.

TABLE III. Systematic uncertainties of the mixing angle between the $h_1(1170)$ and $h_1(1380)$ (in %).

| Source | $b_1(1235)$ | $K_1(1400)$ | $K_1(1270)$ | $h_1(1170)$ | $h_1(1380)$ | Total |
|---|---|---|---|---|---|---|
| Value | 0.7 | 2.1 | 4.2 | 4.1 | 3.6 | 7.2 |

TABLE IV. Branching fractions of $\mathcal{B}(J/\psi \to \eta' h_1(1380)) \times \mathcal{B}(h_1(1380) \to K^*(892)\bar{K} + \text{c.c.})$ and $\mathcal{B}(J/\psi \to \eta' K^*(892)\bar{K} + \text{c.c.})$.

| Source | Branching fraction |
|---|---|
| $\eta' h_1(1380)$ $(\eta' K^+K^-\pi^0)$ | $(1.51 \pm 0.09 \pm 0.21) \times 10^{-4}$ |
| $\eta' h_1(1380)$ $(\eta' K^0_S K^\pm \pi^\mp)$ | $(2.16 \pm 0.12 \pm 0.29) \times 10^{-4}$ |
| $\eta' K^{*\pm}K^\mp$ $(\eta' K^+K^-\pi^0)$ | $(1.50 \pm 0.02 \pm 0.19) \times 10^{-3}$ |
| $\eta' K^{*\pm}K^\mp$ $(\eta' K^0_S K^\pm \pi^\mp)$ | $(1.47 \pm 0.03 \pm 0.17) \times 10^{-3}$ |
| $\eta' K^{*0}\bar{K}^0$ + c.c. $(\eta' K^0_S K^\pm \pi^\mp)$ | $(1.66 \pm 0.03 \pm 0.21) \times 10^{-3}$ |

The uncertainty stems from the total mass uncertainty of the $h_1(1380)$ and the uncertainties from the masses of the other particles, as summarized in Table III. This result is consistent with the ideal decoupling angle 35.26° [11] and theoretical expectations of $(32.3 \pm 1.0)°$ or $(38.3 \pm 1.0)°$ by the Hadron Spectrum Collaboration [30].

## VIII. SUMMARY

In summary, based on a sample of $1.31 \times 10^9$ $J/\psi$ events collected by the BESIII experiment, we report the first observation of $J/\psi \to \eta' h_1(1380)$, where $h_1(1380) \to K^*(892)\bar{K}$ + c.c.. The mass and width of the $h_1(1380)$ are determined to be $M = (1423.2 \pm 2.1 \pm 7.3)$ MeV/$c^2$ and $\Gamma = (90.3 \pm 9.8 \pm 17.5)$ MeV, where the uncertainty from the interference is not included. This measurement is consistent with the previous measurements by the LASS, Crystal Barrel and BESIII Collaborations [2,3,13] with improved precision. The product branching fractions of $h_1(1380)$ production and three body decays are also measured, as shown in Table IV, and isospin symmetry violation is found in $h_1(1380)$ decays between $h_1(1380) \to K^*(892)^+K^-$ + c.c. and $h_1(1380) \to K^*(892)^0\bar{K}^0$ + c.c.. Additionally, based on the measured $h_1(1380)$ mass, the mixing angle between the $h_1(1170)$ and $h_1(1380)$ is determined to be $(35.9 \pm 2.6)°$ assuming the preferred mixing angle between the $K_{1A}$ and $K_{1B}$ of 34°. The measured mixing angle supports the hypothesis that the quark contents of the $h_1(1380)$ is predominantly $s\bar{s}$ and that of the $h_1(1170)$ is predominantly $u\bar{u} + d\bar{d}$.

## ACKNOWLEDGMENTS

The BESIII Collaboration thanks the staff of BEPCII, the IHEP computing center and the supercomputing center of USTC for their strong support. This work is supported in part by National Key Basic Research Program of China under Contract No. 2015CB856700; National Natural Science Foundation of China (NSFC) under Contracts No. 11125525, No. 11235011, No. 11322544, No. 11335008, No. 11425524, No. 11625523, No. 11635010, No. 11375170, No. 11275189, No. 11475164, No. 11475169, No. 11605196, No. 11605198, No. 11705192, No. 11735014; the Chinese Academy of Sciences (CAS) Large-Scale Scientific Facility Program; the CAS Center for Excellence in Particle Physics (CCEPP); the Collaborative Innovation Center for Particles and Interactions (CICPI); Joint Large-Scale Scientific Facility Funds of the NSFC and CAS under Contracts No. U1232201, No. U1332201, No. U1532257, No. U1532258, No. U1532102, No. U1732263; CAS under Contracts No. KJCX2-YW-N29, No. KJCX2-YW-N45, No. QYZDJ-SSW-SLH003; 100 Talents Program of CAS; National 1000 Talents Program of China; Institute of Nuclear and Particle Physics, Astronomy and Cosmology (INPAC) and Shanghai Key Laboratory for Particle Physics and Cosmology; German Research Foundation DFG under Contracts Nos. Collaborative Research Center CRC 1044, FOR 2359; Istituto Nazionale di Fisica Nucleare, Italy; Joint Large-Scale Scientific Facility Funds of the NSFC and CAS; Koninklijke Nederlandse Akademie van Wetenschappen (KNAW) under Contract No. 530-4CDP03; Ministry of Development of Turkey under Contract No. DPT2006K-120470; National Natural Science Foundation of China (NSFC); National Science and Technology fund; The Swedish Resarch Council; U.S. Department of Energy under Contracts No. DE-FG02-05ER41374, No. DE-SC-0010118, No. DE-SC-0010504, No. DE-SC-0012069; U.S. National Science Foundation; University of Groningen (RuG) and the Helmholtzzentrum fuer Schwerionenforschung GmbH (GSI), Darmstadt; WCU Program of National Research Foundation of Korea under Contract No. R32-2008-000-10155-0.